\documentclass{emulateapj}
\usepackage{apjfonts,natbib,epsfig,color}
\usepackage{amsmath} 

\usepackage{graphicx}
\newcommand{\beq}{\begin{equation}}
\newcommand{\beqa}{\begin{eqnarray}}
		  \newcommand{\eeq}{\end{equation}}
\newcommand{\eeqa}{\end{eqnarray}}

\newcommand{\lsim}{\lesssim}
\newcommand{\gsim}{\gtrsim}

\newcommand{\lmk}{\left(}
\newcommand{\rmk}{\right)}

\newcommand{\lkk}{\left[} 
\newcommand{\rkk}{\right]}

\newcommand{\ts}{t_{\rm S}}

\shorttitle{}
\shortauthors{}

\begin{document}


\title{Possibility of a coordinated signaling scheme in the Galaxy and   SETI experiments}


\author{Naoki Seto}
\affil{Department of Physics, Kyoto University, Kyoto 606-8502, Japan}


\begin{abstract}

We discuss a Galaxy-wide coordinated signaling scheme with which a SETI observer needs to examine a tiny fraction of the sky.  The target sky direction is determined as a function of time,  based on high-precision measurements of a progenitor of a conspicuous astronomical event such as a coalescence of a double neutron star binary.    In various respects, such a coordinated scheme would be advantageous for both transmitters and receivers,
and  might be widely prevailing as a tacit adjustment.  For this scheme, the planned space gravitational-wave detector LISA and its follow-on missions have a potential to narrow down the target sky area by a factor of  $10^{3\textit{-}4}$, and could have a large impact on future SETI experiments.

\end{abstract}





\section{introduction}

Since the pioneering work by Drake (1961),  significant efforts have been made for SETI experiments, but no definite signal has been detected so far (see e.g., Tarter et al. 2001;  Siemion et al. 2013; Harp et al. 2016).\footnote{See also https://breakthroughinitiatives.org/initiative/1.}  This might be partly due to the fact that we have explored an extremely small fraction in the multi-dimensional phase space for SETI experiments, under the restrictions of available observational facilities and computational resources (Tarter et al. 2010; Wright et al. 2018). For example, Tarter et al. (2010) figuratively claimed that we have examined only a glass of water randomly sampled from Earth\rq{}s oceans to find a fish (see also Wright et al. 2018).  

In this paper, we discuss a search for intentionally emitted beamed signal from extra-terrestrial intelligence (ETI) in our Galaxy.  There are on the order of $10^{11}$ stars in our Galaxy and many of them are expected to have planets.  While nearby stars are distributed almost isotropically, distant ones ($\gsim 1$kpc) are concentrated around the Galactic plane in the sky. 
When searching for artificial signals from ETIs, we might need to carefully arrange the survey directions as a function of time.  For example, a civilization might use its orbital positions (e.g., periastron position,  planetary eclipse) around the central star to set a natural time window for searchers (Pace \& Walker 1975; McLaughlin 1977; Kipping \& Teachey 2016, see also Makovetskii 1980; Lemarchand 1994; Corbet 1999). 
This kind of timing adjustment of individual senders might be effective, especially for short distance communications. But, if   somewhat distant searchers observe planetary systems densely around the Galactic plane, the incoherent timing windows and cumbersome requirements for the detailed prior investigations (e.g., eclipse epochs)  would significantly complicate the survey.

Considering that  the supposed senders are intelligent, some of them might use a Galaxy-wide signaling scheme with which any searcher needs to examine a tiny fraction of the sky area, as a function of time. Such a sophisticated scheme would be advantageous both for senders and receivers. For example, a sender can significantly suppress the costs for signal transmissions, including the total output power. On another front, a receiver can drastically compactify the SETI phase space, resulting in a reduction of observational facilities and computational resources. Given the reciprocal advantages, a coordinated signaling scheme might be actually prevailing, as a tacit adjustment between involved parties, namely as the Schelling point in the game theory (Schelling 1960; Wright 2018).

In this paper, we first discuss how to configure a  coordinated signaling scheme, without prior communications, but utilizing a conspicuous Galactic event (\S 2).  Then, in \S 3, we point out that  coalescence of  a double neutron star binary (DNSB) would be an attractive candidate for the reference Galactic event. This is partly because the progenitor of the merger event (namely a DNSB) would be suitable for a high-precision measurement with space gravitational wave detectors and radio telescopes, well before the occurrence epoch of the event. 
After the multi-messenger observation of GW170817 (Abbott et al. 2017),   the estimated merger rate of Galactic  DNSBs is also within a preferable range.
With the space gravitational wave detector LISA, the survey sky area might be narrowed down by a factor of $10^{3}$, potentially increasing the prospects for SETI relative to the searches for signatures of primitive life (Lingam \& Loeb  2019).   
Discovery of radio pulsars in a short period DNSB ($\lsim 600$ sec) would be also beneficial for the sky area restriction, and might be realized in an earlier time than LISA.

\section{Coordinated communication scheme}

In this section, we discuss a simple coordinated  communication scheme.  In  this scheme,  we use a conspicuous Galactic astronomical event whose time of the occurrence and the spatial position can be accurately estimated ahead of time, by appropriate observation of its progenitor.  

\subsection{sending cone}

Let us consider a sender S and the progenitor of the reference event E at a Galactic distance $l$  (see Fig.1). For simplicity, we temporarily ignore the motions of both the sender and the progenitor. Also ignoring fluctuations of the metric, we introduce a rest frame covering the Galactic scale.  
Similar to  Nishino \& Seto (2018), we define the time $t_{\rm S,E}$ when the sender S observes the event E with a messenger propagating at the speed of light (e.g.,   electromagnetic wave or gravitational wave). We  also introduce the relative time $\Delta \ts\equiv t-t_{\rm S,E}$ for describing the scheduled actions of the sender.

During the time interval $-2l/c\le \Delta t_{\rm S}\le 0$, the sender S can take the opening angle $\theta$ for transmitting its artificial photon signal, as the solution to the equation below
\beq
\Delta t_{\rm S}=-\frac{l}{c} (1+\cos\theta). \label{th}
\eeq
In the following, we call the time-dependent sending direction as \lq\lq{}sending cone\rq\rq{} (see Fig.1).  It can be regarded as a ring on the celestial sphere.  
Note that the sending duration $2l/c$ is comparable to the light-crossing time of the  Galaxy, namely the time required for a Galactic-scale signal transfer.

With the choice (1) for the sending cone, an artificial signal from S (on the black solid line in Fig.1) will arrive the purple  dashed-circle in Fig.1 at the occurrence time  of the event E.
With this condition,  a receiver $\rm R_\infty$ at the infinity distance catches the S\rq{}s artificial signal simultaneously with the event E. \footnote{This argument is valid for the angle $\rm \angle ESR_\infty <90^\circ$.  But we can easily confirm that eq.(1) holds also for $\rm \angle ESR_\infty >90^\circ$.}  More specifically, for  any sender, the synchronization at infinity can be realized, if and only if its signal is transmitted towards the sending cone determined by eq.(1).

Note that the sending cone moves from $\theta =0$ (at $\Delta t_{\rm S}=-2l/c$) to $\theta =\pi$ (at $\Delta t_{\rm S}=0$), sweeping the whole $4\pi$-sky area at the constant rate. Namely, we have $ (d\cos\theta/d\Delta t_{\rm S})=const$.  In  this sense, the present scheme does not have a preferred sending direction.\footnote{If the duration of the civilization is longer than $2l/c$.  } The center of the sending cone is directed to E for $-2l/c\le \Delta t_{\rm S}<-l/c$, but antipodal to E for $-l/c\le \Delta t_{\rm S}\le 0$.

The transmission scheme (1) was discussed in Nishino \& Seto (2018) for a receiver $\rm R_\infty$ at an extra-Galactic distance (e.g., 40Mpc to NGC4993) much larger than  the Galactic scale $l=O(10)$ kpc, aiming a signal delivery almost synchronously  with the  event E. 
In this paper, we continue to apply the scheme (1),  for a largely different situation, namely intra-Galactic communication with the distance between the sender (S) and receiver ($\rm R_G$) comparable to the Galactic distance $l$ (see Fig.1).   In  this case,  the receiver $\rm R_G$ observes the artificial signal from S  earlier than the event E, with the arrival time  difference of $O(l/c)$.  In fact, the present communication scheme no longer aims a synchronous signal delivery, and, correspondingly,  a sender does not need to care about the distance to a potential receiver (the separation S$\rm R_G$ in Fig.1).

\subsection{receiving cone}

Next we discuss the coordinated signaling scheme from the perspective of a receiver.     For illustrative purpose, we tentatively suppose that the sender S in Fig.1 also attempts to receive an artificial photon signal emitted by  an unknown civilization $\rm S_1$ that follows the Galaxy-wide scheme (1).   Below, for a while, we denote the receiver by S(R) to express its dual nature (sender/receiver). As we mentioned in the previous subsection, the sending cone of the unknown sender $\rm S_1$ sweeps the whole $4\pi$-sky, including the $\rm S_1\to S(R)$ direction. 

The question here is which sky direction the receiver S(R) should search an artificial signal as a function of time. In reality, this is a simple problem, considering the possible propagation direction of the incoming signal (the green arrow in Fig.1).  At any time, the propagation direction must be in the sending cone of the receiver S(R). Otherwise the signal is not synchronized as infinity, and this contradicts with the signaling scheme originally at the unknown sender $\rm S_1$.   
Therefore, the receiver S(R) can limit the survey towards the antipodal direction of its sending cone, reduced from the whole $4\pi$-sky area. 
More specifically, with respect to the reference event E, the opening angle $\theta\rq{}$ of the receiving direction (hereafter \lq\lq{}receiving cone\rq\rq{}) is given by $\theta\rq{}=\pi-\theta $ with the solution $\theta$ for eq.(1).
Of course, the sky direction $\theta\rq{}$ is not changed, even if the receiver S(R) merely searches for other Galactic civilizations, without sending its own signal. Therefore, we hereafter denote a receiver simply by R.

In Fig.2, we show  the geometrical relations between the sending and receiving cones. 
We should notice that, with the present scheme, a receiver can handle a specific sky direction at a time, irrespective of  the distance to a  sender.  The inverse is true for a sender, as mentioned at the end of the previous subsection.    
Considering the simplicities and merits both for the senders and receivers, this signaling scheme (or similar ones) might be actually prevailing as a tacit adjustment in the strategy space  (Schelling 1960; Wright 2018).

\begin{figure}[t]
 \begin{center}
  \includegraphics[width=53mm,angle=270]{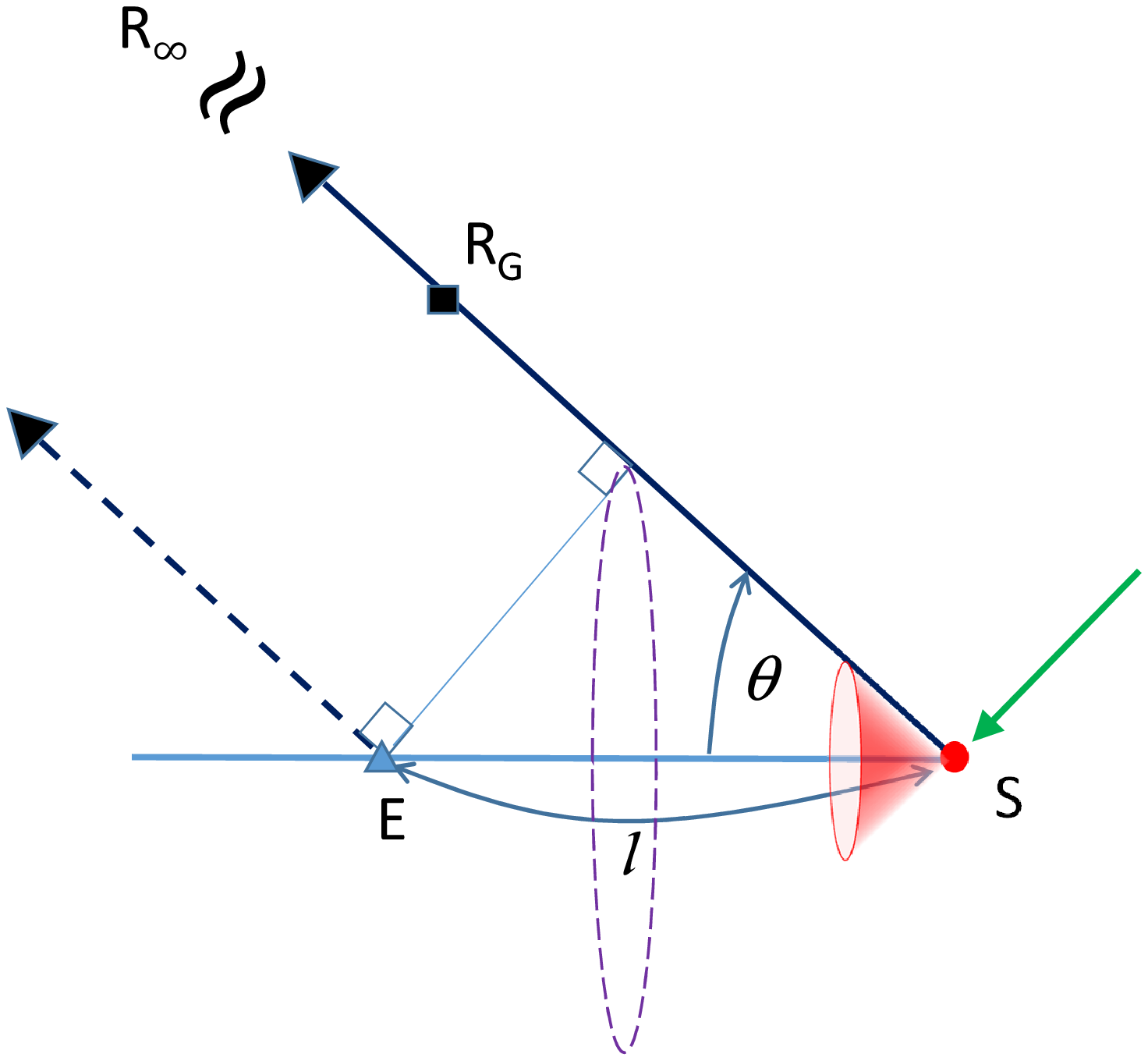}
   \caption{
Our signal transmission scheme using a Galactic reference event E. By observing the progenitor of the reference event E, the sender S estimates the occurrence epoch of E and its position (including the distance $l$).  The sender S transmits its artificial signal to its (red colored) \lq\lq{}sending cone\rq\rq{}  with the time-dependent opening angle $\theta$ given by eq.(1). The artificial signal reaches the purple dashed-circle at the occurrence epoch of the event E.  Then, the receiver $\rm R_\infty$ at infinity catches the artificial signal simultaneously with the event signal.  But a  Galactic receiver $\rm R_G$ observes  the event E later. The green arrow is used in \S 2.2 where S is assumed to be also a receiver and temporarily denoted by S(R). }
  \label{figure:fig1}
 \end{center}
\end{figure}

\begin{figure}
 \begin{center}
  \includegraphics[width=55mm,angle=270]{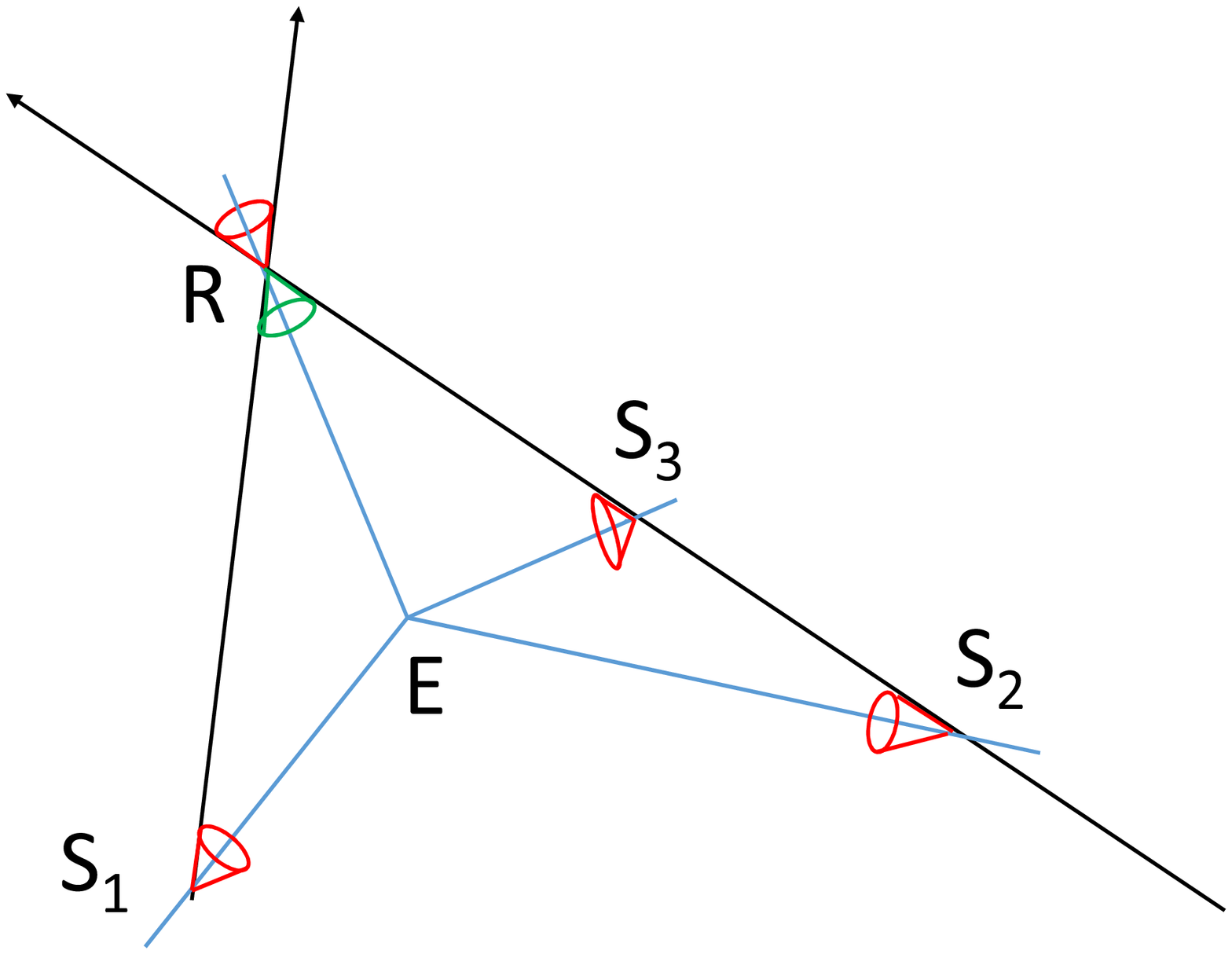}
   \caption{{ The sending and receiving  cones for the coordinated signaling scheme. The sender $\rm S_1$ transmits its artificial photon signal to its time-varying  sending cone (red colored).  Then, at the position of the receiver R, this photon  must be on the sending cone of R, for consistently satisfying the synchronous condition at infinity. In general, once a sender transmits a photon to its sending cone, it will be subsequently on the corresponding sending cone at any future space-time point (e.g., $\rm S_1\to R$ and $\rm S_2\to S_3\to R$). A receiver R only needs to search for its time-varying receiving cone (shown with green color) antipodal to its sending cone. 
} }
  \label{figure:fig1}
 \end{center}
\end{figure}

 \subsection{practical effects}

In reality, the sending/receiving cones have a finite width $\delta \theta$ around the opening angle $\theta$ due to the parameter estimation errors for the progenitor of the event E. If the sky position of the event E can be specified sufficiently well (as expected for typical electromagnetic wave observations), the error width is roughly given as
\beq
\delta \theta\sim \max\lkk \frac{\delta l}l, \frac{\delta (\Delta \ts)}{ (\Delta \ts)} \rkk
\eeq
with the distance error $\delta l$ and the timing error  $\delta (\Delta \ts)$.

Those involved in the communication (e.g., $\rm S_i$ and R in Fig.2) also need to appropriately correct own peculiar motions, including the mean Galactic rotation and random components. Given its measurability, this would not be a serious concern. Meanwhile, a small peculiar velocity would be preferred for the progenitor of E.

Below, when considering a signal transmission from another civilizational to the Earth, we assume that the former has more advanced technology and the associated error width $\delta\theta$ is dominated by our estimation errors given by eq.(2).

 \subsection{Galactic plane}

We now discuss the impact of the Galaxy-wide coordinated signaling scheme for our Galactic  SETI experiments. For comparison, we also study the case for an uncoordinated signaling. We assume that,  except for the time dependence of the sending direction,  there is no difference between the two cases.  For example,   the Galactic structure (e.g., its plane) would be similarly respected for choosing the weight of the  sending directions.   Then, at any incident, the expected numbers of transmitters whose  beams are pointing to us would be similar for the two cases (see Fig. 3). Here we should recall that the senders do not know the positions of receivers including the Earth.  Conveniently, in the coordinated case, we only need to search the receiving cone with the width $\delta \theta$, by automatically excluding almost all the sky directions from the beginning.  Practically, in  view of  the Galactic structure, we will be able to further  limit our survey to the intersections between the region around the Galactic plane and the receiving cone.  Since the information of the Galactic structure is common to the two cases,  the SETI phase space would be compressed by a factor of $\delta \theta$ for the coordinated case, relative to the uncoordinated one.

\begin{figure}
 \begin{center}
  \includegraphics[width=68mm,angle=270]{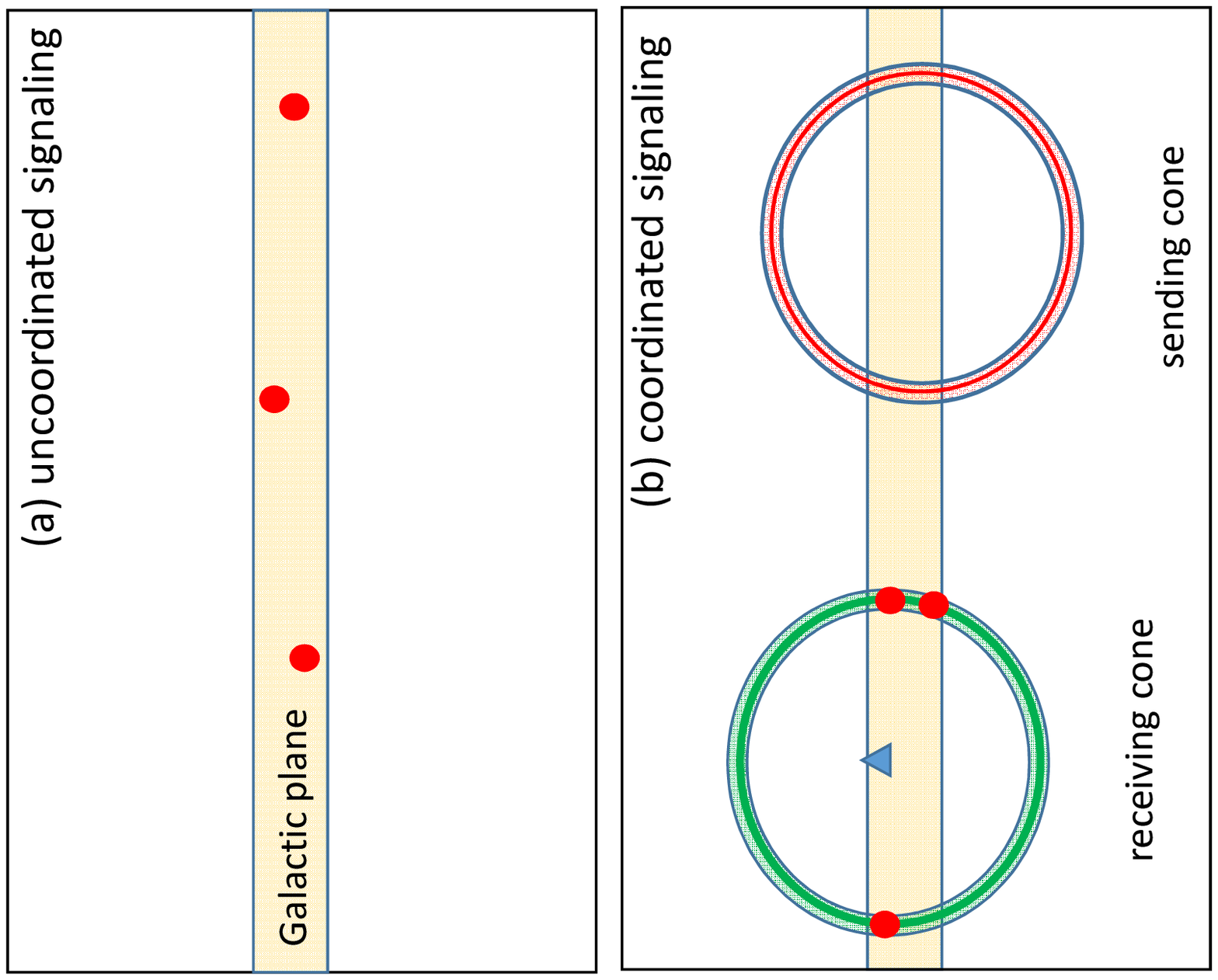}
   \caption{{ The differences between (a) the uncoordinated (random) signaling scheme  and (b) the coordinated scheme,  in the Galactic coordinate. The filled small red circles show the transmitters directed to us at a certain moment. Their numbers would be similar in the two cases.   But, with the coordinated signaling scheme, we can limit the survey direction on the receiving cone (the green ring) whose center is occupied  by the reference event (the blue triangle for $-2l/c\le \Delta t_{\rm S}<-l/c$). The ring has a finite width $\delta \theta$ due to our parameter estimation errors for the reference event.  Our sending cone (the red ring) is antipodal to the receiving cone.  
} }
  \label{figure:fig1}
 \end{center}
\end{figure}

\section{reference Galactic event}
Next we discuss the basic properties that are required  for the reference Galactic event, and propose a DNSB merger as an attractive candidate. This choice might be somewhat affected by the research background of the author and other possibilities would be worth exploring.   In this section, we set  $l=10$kpc as the fiducial value.

\subsection{desired properties}

In order to  less ambiguously and continually select a limited number of the Galactic reference events, its Galactic rate $r_{\rm E}$  should be larger than the inverse of the sending/receiving duration $2l/c\sim 6\times 10^4$yr. But the rate  should  not be overwhelmingly  larger, since the compression factor $\delta \theta$ for the sky area would be  degraded by the total number of equivalent references $2r_{\rm E}l/c$. Namely, the preferable  rate would be roughly given by 
\beq
r_{\rm E}\sim (1\textit{-}10) \times c/2l\sim 1.6\times 10^{-(4\textit{-}5)}\ {\rm yr^{-1}}.
\eeq

Meanwhile, for a Galaxy-wide reference, the progenitor of the event should be easily identified as early as $\sim2l/c$ before the event, at the distance of $l$.  Additionally, its position (especially the distance $l$) and the occurrence epoch $t_{\rm S,E}$ should be estimated at high precision, to reduce the width $\delta \theta$ of the cones.

\subsection{double neutron star merger}

Below, from the viewpoint of mankind (as the only-known example of a 
Galactic civilization),   we argue that a DNSB merger would be suitable for the reference Galactic event.

After the historical multi-messenger observation of GW170817,  the  comoving merger rate of DNSBs is estimated to be $\rm 1540^{+3200}_{-1220}Gpc^{-3}yr^{-1}$ (Abbott et al. 2017).
Using the typical value $\rm 10^{-2}Mpc^{-3}$ for the number density of the Milky-way equivalent galaxies,   we obtain the median value of the  Galactic DNSB merger rate $r_{\rm DNSB}\sim 1.5\times 10^{-4}{\rm yr^{-1}}$.  Therefore, the rate $r_{\rm DNSB}$ is within the preferable range (3).

At $2l/c\sim6\times 10^4$yr before the merger, a DNSB has orbital period of $\sim 600$ sec.  A recycled pulsar  in such a short-period orbit would be an interesting target for radio telescopes such as SKA (Dewdney et al. 2009).  But, due to a significant Doppler smearing, it would  not be straightforward to find a  recycled pulsar by taking a Fourier transformation of the radio data (see e.g., Lorimer \& Kramer 2004).  
The smearing effect would not be severe for the younger (normal) pulsar in a DNSB, and subsequent identification of the recycled one would allow us to make a desired high-precision measurement. However, the younger one is expected have a smaller beaming fraction (Levin et al. 2013) and would typically have a shorter lifetime before crossing the death line. 

Fortunately, the DNSB emits strong a gravitational wave signal at $\sim 3$mHz, and would be a promising target for the Laser Interferometer Space Antenna (LISA. Amaro-Seoane et al. 2017) and TianQin (Luo et al. 2016).
Indeed, taking a long integration time $\sim 10$yr and resolving Galactic foreground, LISA would detect the binary at a distance of $l\sim 10$kpc with the signal-to-noise ratio of $\rm SNR\sim 500$ (Takahashi \& Seto 2002). The subsequent follow-up radio observation for a  recycled pulsar would be much easier than a blind search (Kyutoku et al. 2019). If succeeded, the radio data provide us with the sky position, the residual eccentricity and the orbital inclination much better than LISA data alone. Then the distance $l$ (and correspondingly the cone width $\delta \theta$) would be estimated at $\delta \theta \sim \delta l/l\sim  {\rm SNR}^{-1}\sim 2\times 10^{-3}$ with a negligible contribution of the timing error in eq.(2). 
For a DNSB at $l\lsim 3$kpc, we might obtain a better distance estimation by a  long-term parallax measurement with a radio telescope  (Smits et al. 2009). Furthermore, with the next-generation space missions such as DECIGO (Kawamura et al. 2011), we would reach $\delta \theta \sim \delta l/l\sim 10^{-4}$, depending on the integration time. 

Finally, we  should mention that the burst-like gravitational and electromagnetic waves around the merger are not directly relevant for the present transmission scheme.  This is different from the extra-Galactic transmission for which the actual synchronization would be crucial  (Nishino \& Seto 2018).   But, if we have the luck to discover a Galactic  DNSB that will merge shortly (e.g. in $\sim 100~$yr), it might be interesting to consider the possibility of synchronous signal delivery.

\section{discussion}

In this paper, we discussed a coordinated signaling scheme in our Galaxy.  This scheme would be advantageous both for senders and receivers, in various respects. However, not merely the possibility of their intentional signaling, but even the existence of other Galactic ETIs is totally unclear at present. Based on our  social-scientific standpoint, it might seem reasonable that the adopted signaling scheme would depend on the age of the civilization and their knowledge on other Galactic civilizations obtained from their past SETI-like activities.  But, given the  large uncertainties,  further detailed presumptions about
the likely forms of the signaling scheme would be far beyond the scope of this paper.

If all the signaling Galactic civilizations follow the present coordinated scheme (as an extreme case), we need the time interval $2l/c\sim6\times10^4$yr to complete a single scan of the Galaxy. This is  comparable to the light-crossing time in our Galaxy, namely the typical time for Galactic communication. Unfortunately, it is much longer than our individual lifetime.  In  this relation, if we have  totally $N_{\rm T}$ signaling ETIs in our Galaxy, the expected number $n_{\rm S}$ of the scanned ones would be
\beq
n_{\rm S}\sim N_T\lmk \frac{2l}c  \rmk^{-1}\Delta T\sim 1 \lmk \frac{N_{\rm T}}{2000}  \rmk\lmk \frac{2l/c}{\rm 6\times 10^4 yr}  \rmk^{-1} \lmk \frac{\Delta T}{\rm 30yr}  \rmk 
\eeq
within the observational times $\Delta T$.  The total number $N_{\rm T}$ is quite uncertain and could be much smaller than 2000.  Therefore, in reality, a practical approach would be a mixed strategy, assigning some faction of survey time to the specific directions associated with the receiving cone, in addition to the traditional random pointing.    If we succeed to detect a civilization in our receiving cone within the time span $\Delta T$,  the total number of the Galactic  civilizations following the signaling scheme is roughly estimated to be $\sim 2l/(c\Delta T)$.

\acknowledgments
 This work is supported
by JSPS Kakenhi Grant-in-Aid for Scientific Research
(Nos. 15K65075, 17H06358).

\end{document}